
\vsize=8.6truein
\hsize=6.0truein
\hoffset=.25in
\voffset=.25in

\tolerance 500
\hfuzz=50pt


\font\twelverm = cmr12
\font\twelvei  = cmmi12
\font\twelvesy = cmsy10 at 12pt 
\font\twelvebf = cmbx12
\font\twelveit = cmti12
\font\twelvesl = cmsl12
\font\twelvett = cmtt12

\font\tenex = cmex10

\font\ninerm =  cmr9
\font\ninei  = cmmi9
\font\ninesy = cmsy9
\font\ninebf = cmbx9

\font\sixrm=cmr6
\font\sixi =cmmi6
\font\sixsy=cmsy6
\font\sixbf=cmbx6


\def\twelvept{\def\rm{\fam0\twelverm}
  \textfont0=\twelverm   \scriptfont0=\ninerm    \scriptscriptfont0=\sixrm
  \textfont1=\twelvei    \scriptfont1=\ninei     \scriptscriptfont1=\sixi
  \textfont2=\twelvesy   \scriptfont2=\ninesy    \scriptscriptfont2=\sixsy
  \textfont3=\tenex      \scriptfont3=\tenex     \scriptscriptfont3=\tenex
  \textfont\bffam=\twelvebf   \scriptfont\bffam=\ninebf
  \scriptscriptfont\bffam=\sixbf  \def\bf{\fam\bffam\twelvebf}
  \textfont\itfam=\twelveit       \def\it{\fam\itfam\twelveit}
  \textfont\slfam=\twelvesl       \def\sl{\fam\slfam\twelvesl}
  \textfont\ttfam=\twelvett       \def\tt{\fam\ttfam\twelvett}
  \normalbaselineskip=16pt plus 1pt
  \setbox\strutbox=\hbox{\vrule height11pt depth5pt width0pt}
  \let\sc=\ninerm
  \normalbaselines\rm}

\def\footnoterule{\kern-3pt \hrule width \hsize \kern2.6pt}
\def\square{\kern1pt\vbox{\hrule height1.2pt\hbox{\vrule width1.2pt\hskip3pt
\vbox{\vskip 6pt}\hskip 3pt\vrule width.6pt}\hrule height.6pt}\kern1pt}
\def\diag{\,{\rm diag}\,}

\baselineskip 12pt plus 1pt minus 1pt
\vskip 12pt
\centerline{{\bf GAUGE THEORIES FOR LINEAL GRAVITIES}\footnote{*}
{\ninerm This work is supported in part by funds provided by the U. S.
Department of Energy (D.O.E.) under contract \#DE-AC02-76ER03069.}}
\vskip 20pt
\centerline{R. JACKIW}
\vskip 5pt
\centerline{\it Center for Theoretical Physics}
\centerline{\it Laboratory for Nuclear Science}
\centerline{\it and Department of Physics}
\centerline{\it Massachusetts Institute of Technology}
\centerline{\it Cambridge, Massachusetts\ \ 02139\ \ \ U.S.A.}
\vskip .4in

\centerline{ABSTRACT}
\smallskip
{\narrower
Here is summarized the gauge theoretical formulation and quantization
of two  popular gravity theories in (1+1)-dimensional time.
\smallskip}

\twelvept

\vskip16pt
\baselineskip 14pt plus .5pt minus .5pt
\line {\bf 1.\quad INTRODUCTION \hfil}
\vskip 8pt

We study lower-dimensional gravity both for pedagogical reasons --- one
expects that the dimensional reduction effects sufficient simplification to
permit thorough analysis, while still retaining useful content to inform the
physical $(3+1)$-dimensional problem --- and also, if one is lucky, there are
practical applications --- {\it e.g.\/}
idealized cosmic strings are described by
$(2+1)$-dimensional gravity, while the still lower-dimensional models are used
in statistical mechanics.

The drastic dimensional reduction to $(1+1)$ dimensions --- gravity on a line,
{\it i.e.\/}, {\it lineal\/} gravity --- is not devoid of interest, provided
dynamical equations are not based on the Einstein tensor $G_{\mu\nu} =
R_{\mu\nu} - {1\over 2} g_{\mu\nu} R$, which vanishes identically in two
dimensions.

In a proposal of several years ago,$^1$
it was suggested that gravity equations be
based on the Riemann scalar $R$, the simplest entity that encodes
in two dimensions all local geometric information about space-time.
Moreover, in an action formulation it is necessary to introduce an additional
scalar field, which acts as a Lagrange multiplier that enforces the equation of
motion for $R$.  Thus we are dealing with scalar-tensor theories, or --- to
use the contemporary string nomenclature --- ``dilaton'' gravities.

Since the initial proposal, various models have been studied.  Here I shall
describe two that are selected by their group theoretical properties:
they can be formulated as gauge theories based on groups relevant to
space-time: de~Sitter or anti-de~Sitter (in $(1+1)$-dimensions
both groups are $SO(2,1)$, although the geometries are different)
and Poincar\'e.  The first of these is
the one proposed originally;$^1$  it is governed by the action
$$I_1 = \int d^2x\,\sqrt{-g}\, \eta (R-\Lambda)\eqno(1)$$
The second is ``string-inspired'' and has been recently studied for purposes
of modeling (on a line!) black hole physics;$^2$ its action is
$$\bar{I}_2 = \int d^2x\,\sqrt{-\bar{g}}\, e^{-2\varphi}\left( \bar{R} + 4
\bar{g}^{\mu\nu} \partial_\mu\varphi\partial_\nu \varphi - \Lambda\right)
\eqno(\overline{2})$$
(Notation: time and space carry the metric tensor
${\bar{g}}_{\mu\nu}$ with signature $(1,-1)$.  The two-vector $x^\mu = (t,x)$
will be
frequently presented in light-cone components $x^\pm \equiv {1\over
\sqrt{2}}{(t\pm x)}$.  Tangent space components are labeled by Latin letters
$a,b,\ldots$, and the Minkowski metric tensor $h_{ab}=\diag (1,-1)$
raises/lowers these indices.  Also we use the anti-symmetric tensor
$\epsilon^{ab}$, $\epsilon^{01} = 1$.)

In (1), $R$ is the scalar curvature built from $g_{\mu\nu}$, $\eta$ is a
world scalar Lagrange multiplier related to the dilaton, while $\Lambda$ is a
cosmological constant.  In ($\overline{2}$)
we temporarily use an over-bar to denote a
differently scaled metric tensor $\bar{g}_{\mu\nu}$ from which $\bar{R}$ is
constructed, while $\varphi$ is the dilaton.  Formula ($\overline{2}$)
arises naturally
from string theory, restricted to a two-dimensional target space, with the
anti-symmetric tensor field identically vanishing.
In the string context, matter is
taken to couple to $\bar{g}_{\mu\nu}$; for our purposes in the absence of
matter it is convenient to redefine variables by $\bar{g}_{\mu\nu}=
e^{2\varphi} g_{\mu\nu}$, $\eta = e^{-2\varphi}$.  Then ($\overline{2}$)
becomes
$$I_2 = \int d^2x\, \sqrt{-g}\, \left( \eta R - \Lambda\right) \eqno(2)$$
but it is to be remembered that because of the redefinition, the ``physical''
metric tensor is $g_{\mu\nu}/{\eta}$.  Note that (2) is invariant against
shifting $\eta$ by a constant, because $\sqrt{-g}\,R$ is a total derivative.

It is seen that the two models (1) and (2) differ in the placement of the
Lagrange multiplier with the cosmological term:  in (1) $\eta$ multiplies
$\Lambda$, in (2) the $\eta$ factor is absent from $\Lambda$.
Of course in the limit $\Lambda=0$, the difference disappears.

We now describe the interesting gauge group structure of (1) and (2) which we
name {\it (anti)~de~Sitter gravity\/} and {\it extended Poincar\'e gravity\/},
respectively.
\goodbreak
\bigskip
\noindent{\bf II.\quad (ANTI)~DE~SITTER GRAVITY}
\medskip
\nobreak
The equations of motion that follow from varying $\eta$ and $g_{\mu\nu}$ in (1)
are
$$R=\Lambda\eqno(3)$$
$$\left( {\cal D}_\mu {\cal D}_\nu - g_{\mu\nu} {\cal D}^2\right) \eta -
{\Lambda\over 2} g_{\mu\nu}\eta = 0 \eqno(\hbox{4a})$$
The second equation, with ${\cal D}_\mu$ the space-time covariant derivative,
can be decomposed into traceless and trace parts.
$$\eqalignno{\left( {\cal D}_\mu {\cal D}_\nu - {1\over 2} g_{\mu\nu} {\cal
D}^2\right) \eta &= 0 & (\hbox{4b}) \cr
\left( {\cal D}^2 + \Lambda\right) \eta &= 0 &(\hbox{4c}) \cr}$$

The above geometric dynamics may be presented in a gauge theoretical
fashion.$^3$  To this end one
uses the (anti) de~Sitter group with Lorentz generator $J$ and
translation generators $P_a$ satisfying the $SO(2,1)$ algebra (for
$\Lambda\not=0$).
$$\left[ P_a,J\right] = \epsilon_a{}^b P_b\ \ ,\qquad \left[ P_a, P_b\right] =
{\Lambda\over 2} \epsilon_{ab} J \eqno(5)$$
The gauge connection one-form is introduced $A=A_\mu\,dx^\mu$ and expanded
in terms of the generators,
$$A = e^a P_a+\omega J\eqno(6)$$
where $e^a_\mu$ is the {\it Zweibein\/} and $\omega_\mu$ is the
spin-connection.  The curvature two-form
$$F = dA + A^2 \eqno(7)$$
becomes
$$F = f^a P_a + fJ = \left( De\right)^a P_a + \left( d\omega + {\Lambda\over 4}
e^a \epsilon_{ab} e^b\right) J \eqno(8)$$
$$\left( De\right)^a \equiv de^a + \epsilon^a{}_b \omega e^b\eqno(9)$$
It is seen that $d\omega$ is proportional to
the scalar curvature density and $f^a = \left(De\right)^a$
is proportional to the torsion density,
each expressed in terms of $e^a$ and $\omega$, which at this
stage are independent variables.

The Lagrange density
$$\eqalign{{\cal L}'_1 &= \sum\limits^2_{A=0} \eta_A F^A
= \eta_a \left( De\right)^a +
\eta_2 \left( d\omega + {\Lambda\over 4}e^a \epsilon_{ab} e^b\right) \cr
F^A &= \left( f^a, f\right)\ \ ,\qquad \eta_A = \left( \eta_a,
\eta_2\right)}\eqno(10)$$
is gauge invariant: the three field strengths $F^A$ transform covariantly
according to the three-dimensional adjoint representation, while the
Lagrangian multiplier triplet $\eta_A$ transforms by the coadjoint
representation.

The equation obtained from (10) by varying $\eta_a$ gives the
condition of vanishing torsion, and allows evaluating the spin connection in
terms of the {\it Zweibein\/}.
$$\omega = e^a \left(h_{ab} \epsilon^{\mu\nu} \partial_\mu e^b_\nu\right)\big/
\det e\eqno(11)$$
The equation which follows upon variation of $\eta_2$ regains (3) once (11) is
used.  Variation
of $e^a$ and $\omega$ produces equations for the Lagrange multipliers
$\eta_a$ and $\eta_2$, respectively, the latter coinciding with
$2\eta$ in the geometric formulations (1), (3) and (4).
$$\eqalignno{d\eta_a + \epsilon_a{}^b \omega \eta_b + {\Lambda\over 2}
\epsilon_{ab} \eta_2 e^b &= 0 &(\hbox{12a}) \cr
d\eta_2 + \eta_a \epsilon^a{}_b e^b &= 0 &(\hbox{12b})\cr}$$
Upon taking a space-time
covariant derivative of (12b) and using (12a) to eliminate
$\eta_a$, we recover (4).  Finally we see that when $\omega$ is eliminated from
${\cal L}'_1$ with the help of (11), so that the torsion (9) vanishes,
what remains is the Lagrange density of
(1), expressed in terms of {\it Zweibeine\/}.

Thus the geometric formulation of this gravity theory is contained within the
(anti)~de~Sitter
group theoretical framework for solutions with $\det e\not=0$, but
see below.

Explicit classical solutions to the
equations are easy to find.  Working within the
geometric framework, we use coordinate invariance to choose a conformally
flat metric tensor.
$$g_{\mu\nu} = h_{\mu\nu} \exp 2\sigma \eqno(13)$$
Then (3) becomes the Liouville equation.
$$\square \sigma = -{\Lambda\over 2} \exp 2\sigma \eqno(14)$$
Its general solution depends on two arbitrary functions
of the two light-cone variables, $F(x^+)$, $G(x^-)$,
$$\exp 2\sigma = {F'(x^+) G'(x^-)\over \left( 1 + {\displaystyle{\Lambda\over
4}} FG\right)^2 } \eqno(15)$$
whose derivatives fulfill the consistency condition $F'G'>0$.  But the residual
coordinate invariance within the conformal gauge allows choosing $F(x^+) =
x^+, G(x^-) = x^-$, hence
$$\exp 2\sigma = {1\over \left( 1 + {\displaystyle{\Lambda\over 8}}
x^2\right)^2} \eqno(16)$$
In conformal gauge, (4b) reduces to
$$\partial_\mu V_\nu + \partial_\nu V_\mu - h_{\mu\nu} h^{\alpha\beta}
\partial_\alpha V_\beta =0\eqno(17)$$
where $V_\mu$ is defined by
$$V_\mu \exp 2\sigma=\partial_\mu\eta \eqno(18)$$
Equation (17) is just the (flat-space) conformal Killing equation with
solutions in terms of arbitrary functions of a single light-cone variable.
$$V_- = V_-(x^+)\ \ ,\qquad V_+ = V_+(x^-)\eqno(19)$$
Finally the remaining equation (4c) together with (18) restricts these
functions, so that the solution for $\eta$ takes the form
$$\eta = {\alpha_a x^a + \alpha_2 \left( 1 -
{\displaystyle{\Lambda\over 8}}x^2\right) \over 1 +
{\displaystyle{\Lambda\over 8}} x^2} \eqno(20)$$
where $\alpha_a$ is a constant two-vector and $\alpha_2$ is a constant scalar.

The {\it Zweibein\/} and spin connection of the gauge theoretical formulation
are given by related formulas.  The former, the ``square root'' of the
metric tensor, becomes (apart from an arbitrary Lorentz transformation on the
tangent-space indices)
$$e^a_\mu =  \delta^a_\mu\exp\sigma = {1\over 1 +
{\displaystyle{\Lambda\over 8}} x^2} \delta^a_\mu \eqno(21)$$
while the latter is
$$\omega_\mu = - h_{\mu\alpha}\epsilon^{\alpha\beta} \partial_\beta
\sigma\eqno(22)$$
The Lagrange multiplier $\eta_2$
coincides with $2\eta$, while Eq.~(12) for $\eta_a$ is solved by
$$\eta_a \exp \sigma = 2\epsilon_a{}^\mu \partial_\mu \eta \eqno(23)$$
Of course the general solution is an arbitrary coordinate transformation of
the above.

Finally we observe that the gauge theoretical formulation allows an
alternative group theoretical
presentation of solutions.  The field equations following from
(10), upon respective variation of $\eta_A$ and $A$, are
$$F=0\eqno(24)$$
$$dH + [A,H] = 0 \eqno(25)$$
$A$, $F$ and $H= \eta_a h^{ab} P_b - {\Lambda\over2} \eta_2 J$
belong to the $SO(2,1)$ algebra (the factor $\Lambda / 2$
is a consequence of the group
metric).  Equation (24) implies that $A$ is a pure gauge given
by an arbitrary element $U$ of the $SO(2,1)$ group,
$$A = U^{-1}dU \eqno(26)$$
while the Lagrange multiplier is then determined by (25) to be
$$H = U^{-1} \Phi\,U\eqno(27)$$
where $\Phi$ is a constant element in the algebra.  The explicit group and
algebra elements that correspond to the above solution, Eqs.~(20) -- (23), are
$$
\eqalignno{
U &=
e^{-i \pi J} \,
e^{x^{+} P_{+}} \,
e^{-\ln \left( 1+{\Lambda\over8}x^2 \right) J} \,
e^{x^{-} P_{-}}
&(28) \cr
\noalign{\hbox{and}}
\Phi &={2\over\Lambda} \alpha_a \epsilon^{ab} P_b + \alpha_2 J   &(29) \cr}$$
$U$ is unique up to a constant gauge transformation.
Performing such a gauge transform allows setting two of the three constants
$\left( \alpha_a, \alpha_2 \right)$ to zero, so that
the invariant dependence is on a single quantity,
which we can choose as
$
-{4\over\Lambda^{\!2}} \left( \alpha^a\alpha_a
-{\Lambda\over2}\alpha_2\alpha_2 \right)
= -\eta^a\eta_a + {\Lambda\over2}\eta_2\eta_2\ .
$

Within the gauge theoretical framework, an even simpler solution to (24) and
(25) is available: $A=0$, $H=\Phi$, which makes no sense geometrically: not
only $\det e$, but both the connections $e^a$ and $\omega$ vanish!  But in
fact use can be made of such solutions:
when presented with a geometrically singular configuration,
perform any gauge transformation producing non-singular connections,
for example with the group element $U$ above. So we see that the group
theoretical framework, even in its $\det e=0$ sector, contains adequate
information for encoding the gravity theory.
\goodbreak
\bigskip
\noindent{\bf III.\quad EXTENDED POINCAR\'E GRAVITY}
\medskip
\nobreak
Equations of motion of the string-inspired gravitational theory (2) are,
from varying~$\eta$
$$R = 0 \eqno(30)$$
and from varying $g_{\mu\nu}$
$$\left( {\cal D}_\mu {\cal D}_\nu - g_{\mu\nu} {\cal D}^2\right) \eta -
{\Lambda\over 2}  g_{\mu\nu} = 0 \eqno(\hbox{31a})$$
which is equivalent to
$${\cal D}_\mu {\cal D}_\nu \eta = -{\Lambda\over 2} g_{\mu\nu}
\eqno(\hbox{31b})$$
Note that (31a) differs from (4a) by the absence of $\eta$ in the last term.

To give a gauge theoretical formulation,$^4$ we make use of the {\it centrally
extended\/} Poincar\'e group, whose algebra is
$$\left[ P_a, J\right] = \epsilon_a{}^bP_b\ \ ,\qquad \left[ P_a, P_b\right] =
\epsilon_{ab} I \eqno(32)$$
where the central element $I$ commutes with $P_a$ and $J$.  Consequently the
connection $A$ and curvature $F$ now become
$$\eqalignno{A &= e^a P_a + \omega J + a I &(33) \cr\noalign{\vskip 0.2cm}
F &= dA + A^2 = f^a P_a+ f J+ g I \cr
&=\left( De\right)^a P_a + d\omega J + \left( da + {1\over 2}e^a \epsilon_{ab}
e^b\right) I &(34) \cr}$$
Here $a$ and $g$ are the additional connection and curvature associated with
the central element in the algebra.

This magnetic-like extension of the Poincar\'e group may be viewed as an
unconventional contraction of the de~Sitter group: The ordinary Poincar\'e
algebra (Eq.~(32) without the central element) is the $\Lambda \to 0$
contraction of the $SO(2,1)$ algebra (5).  However, owing to the well-known
ambiguity of two-dimensional angular momentum, in (5) one may replace $J$ by
$J - 2I/\Lambda$ before taking the $\Lambda\to 0$ limit, which then leaves
(32).

The extension reflects a 2-cocycle in the composition law for representatives
of
the Poincar\'e group.  If the group acts on coordinates $x^a$ by
$$x^a\longrightarrow \bar{x}^a = {\cal M}^a{}_b x^b + q^a \eqno\hbox{(35a)}$$
where ${\cal M}$ is a finite Lorentz transformation
$${\cal M}^a{}_b = \delta^a{}_b \cosh \alpha + \epsilon^a{}_b \sinh \alpha
\eqno(\hbox{35b})$$
and $q^a$ is a finite translation, the composition law for these is
$$\eqalignno{{\cal M}_{(12)} &= {\cal M}_{1} {\cal M}_{2} &(\hbox{36a})
\cr
q_{(12)} &= q_{1} + {\cal M}_1 q_{2} &(\hbox{36b})\cr}$$
However, the composition law for a representation $G({\cal M},q)$ containing
the extension (32) in its algebra acquires a 2-cocycle.
$$G\left( {\cal M}_1, q_1\right) G\left( {\cal M}_2, q_2\right) =\exp\left\{
{i\over 2} q^a_1 \epsilon_{ab} \left( {\cal M}_1 q_2\right)^b \right\} G
\left( {\cal M}_1 {\cal M}_2, q_1 + {\cal M}_1 q_2\right) \eqno(37)$$
($I$ is represented by $i=\sqrt{-1}$.)

A finite gauge transformation, generated by the gauge function $\Theta$,
$$\Theta = \theta^a P_a + \alpha J + \beta I \eqno(38)$$
produces the following transformations on the connections.
$$\eqalign{e^a \to \bar{e}^a &= \left( {\cal M}^{-1}\right)^a_{\ b} \left( e^b
+ \epsilon^b{}_c \theta^c \omega + d\theta^b\right) \cr
\omega \to \bar{\omega} &= \omega + d\alpha \cr
a\to \bar{a} &= a - \theta^a \epsilon_{ab} e^b - {1\over 2} \theta^2\omega +
d\beta + {1\over 2} d\theta^a \epsilon_{ab} \theta^b \cr}\eqno(39)$$
The multiplet of curvatures $F^A = \left( f^a, f,g\right)$ transforms by the
adjoint $4\times4$ representation of the extended group,
$$\eqalign{f^a \to \bar{f}^a
&= \left( {\cal M}^{-1}\right)^a_{\ b} \left( f^b +
\epsilon^b{}_c \theta^c f\right) \cr
f \to \bar{f} &= f \cr
g\to \bar{g} &= g -\theta^a \epsilon_{ab} f^b - {1\over 2} \theta^2 f
\cr}\eqno(40)$$
or
$$\eqalign{ F^A \to \bar{F}^A &= \sum\limits^3_{B=0} \left(
U^{-1}\right)^A_{\ B} F^B\cr
U&= \left( \matrix{
{\cal M}^a{}_b & - \epsilon^a{}_c \theta^c & 0 \cr
0 & 1 & 0 \cr
\theta^c \epsilon_{cd} {\cal M}^d{}_b & - \theta^2/2 & 1 \cr}\right)
\cr}\eqno(41)$$
The upper left $3\times3$ block in $U$ comprises the adjoint representation
of the conventional Poincar\'e group with $q^a$ of (35) identified with
$-\epsilon^a{}_c\theta^c$, while the fourth row and column arise from the
extension.  Note that in the above realization of the gauge action on $F$, the
extension is not visible: $I$ is represented by ${\bf O}$.
 On the other hand, an
additional connection and curvature ($a,g$) are present.

In this representation, the extended algebra possesses a non-singular Killing
metric, which is unavailable without the extension.
$$h_{AB} = \left( \matrix{ h_{ab} & \phantom{-}0 & \phantom{-}0 \cr
0 & \phantom{-}0 & -1 \cr
0 & -1 & \phantom{-}0 \cr}\right) \eqno(42)$$
It is true that ${}^TUh U = h$; this allows raising and lowering the indices
$(A,B)$.
Additionally, there exists an invariant four-vector
$$
i^A = \left(\matrix{\phantom{-}0\cr \phantom{-}0\cr -1\cr} \right)\ ,
\ \ i_A = \left( \ 0,\ 1,\ 0\ \right) \eqno(43)
$$
for which it is true that
${\left( U^{-1} \right)}^A\!\phantom{}_B i^B=i^A$
and $i_B U^B{}_{\!A} = i_A$.
[The occurence of such invariant vectors is related to the fact that the
algebra (32) is solvable.]

An invariant Lagrange density is now constructed with an extended multiplet of
Lagrange multipliers $\eta_A$,
$$\eqalign{
{\cal L}'_2 &= \sum\limits^3_{A=0} \eta_A F^A
\ \ =\ \ \eta_a \left( De\right)^a +
\eta_2 d\omega + \eta_3 \left( da + {1\over 2} e^a \epsilon_{ab} e^b\right)\cr
F^A &= \left( f^a, f,g\right)\ \ ,\qquad \eta_A = \left( \eta_a,
\eta_2,\eta_3\right) \cr}\eqno(44)$$
which obey the coadjoint transformation law,
$$\eta_A \to \bar{\eta}_A = \sum^3_{B=0} \eta_B U^B{}_A
\eqno(45)$$
or in components
$$\eqalign{\eta_a \to \bar{\eta}_a &= \left( \eta_b - \eta_3 \epsilon_{bc}
\theta^c\right) {\cal M}^b{}_a \cr
\eta_2 \to \bar{\eta}_2 &= \eta_2 - \eta_a \epsilon^a{}_b \theta^b - {1\over
2}\eta_3 \theta^2 \cr
\eta_3\to\bar{\eta}_3 &= \eta_3 \cr}\eqno(46)$$

Using the invariant metric (42) and the invariant vector (43),
other group invariants may be constructed.
$$
\eqalignno{{\cal F}^2 &= \sum^3_{A,B=0} {}^*F^A h_{AB} F^B &(47a) \cr
M &=-{2\over\Lambda} \sum^3_{A,B=0} \eta_A h^{AB} \eta_B &(47b) \cr
C &= 2 \eta_A i^A &(47c) \cr}
$$
where ${}^*F^A$ is the 0-form ${1\over 2} \epsilon^{\mu\nu} F^A_{\mu\nu}$, dual
to the 2-form $F^A$.

We recognize in (43) the torsion $\left( De\right)^a$ and curvature $d\omega$
densities, which vanish as a consequence of varying $\eta_a$ and $\eta_2$,
respectively.  Thus Eq.~(30) is regained.  The Lagrange multiplier $\eta$ in
(2) corresponds to ${1\over2}\eta_2$ in the present formulas
and the equation for it, obtained by varying
$\omega$, is as in the (anti)~de~Sitter model, (12b),
$$d\eta_2 + \eta_a \epsilon^a{}_b e^b = 0 \eqno(\hbox{48a})$$
while the equation for $\eta_a$, obtained by varying $e^a$, differs from
(12a),
$$d\eta_a + \epsilon_a{}^b \omega \eta_b + \eta_3\epsilon_{ab} e^b=0
\eqno(\hbox{48b})$$
We need a value for $\eta_3$ to close the system (48).  The equation for that
multiplier is obtained by varying $a$,
$$d\eta_3 = 0 \eqno(\hbox{48c})$$
and a constant, cosmological solution
$$\eta_3 = - {\Lambda\over 2}\eqno(\hbox{48d})$$
renders (48b) similar to (12a),
$$d\eta_a + \epsilon_a{}^b \omega \eta_b - {\Lambda\over 2}\epsilon_{ab}
e^b = 0 \eqno(\hbox{48e})$$
except that there is no factor of $\eta_2$ in the last, cosmological term of
(48e).  This of course has the consequence that when (48a) and (48e) are
combined as before, the second order equation that emerges for
$\eta = {1\over2}\eta_2$
reproduces (31).

The remaining equation of the gauge theoretical formulation, obtained by
varying $\eta_3$
$$da = - {1\over 2} e^a \epsilon_{ab} e^b\eqno(49)$$
and allowing evaluation of $a$, has no counterpart in the geometric
formulation.
Equation (49) can always be locally integrated because the right side is
a two-form, hence closed in two dimensions.  However in general, there will be
singularities in $a$, since upon integrating (49) over a two-space, the
right side gives the total ``volume,'' which could be a well-defined
non-vanishing quantity, while the left side always integrates to zero if
the manifold is closed and bounded, and $a$ is non-singular.

Note that upon eliminating $\omega$ in ${\cal L}'_2$ with the zero-torsion
equation $\left( De\right)^a=0$ and evaluating $\eta_3$ at $-\Lambda/2$, ${\cal
L}'_2$ coincides with the Lagrange density in (2), now expressed in terms of
{\it Zweibeine\/}, apart from the total derivative $-\Lambda/2\,da$, which
does not contribute to equations of motion.

Thus here again, the group theoretical formulation reproduces the geometric
one, for solutions with $\det e\not=0$, but again see below.
However, the former is more flexible:
Eq.~(48c) is satisfied with vanishing $\eta_3$; this corresponds to a
vanishing cosmological constant.  Thus the gauge theory built on the {\it
extended\/} Poincar\'e group possesses as a solution a {\it non-extended\/}
system.  It is interesting therefore that here the cosmological term is an
integration constant, and not inserted {\it a priori\/} into the theory.

Finding explicit
solutions is straightforward.  In the geometric formulation, (3) is
solved by a flat metric tensor.
$$g_{\mu\nu} = h_{\mu\nu}\eqno(50)$$
Then (31) immediately gives
$$2\eta= M- {\Lambda\over 2}\left( x - x_0\right)^2  \eqno(51)$$
with $M$ and $x_0$ being integration constants,
the former reflecting the $\eta$-translation invariance mentioned earlier.

Interest in the model$^2$
derives precisely from the above ``black-hole'' solution with mass $M$ [in
terms of the ``physical'' metric $g_{\mu\nu}/{\eta}$], located at $x_0$.
An arbitrary coordinate transformation of this configuration produces the
general solution.

The gauge theoretical counterparts of the above are a flat {\it Zweibein\/}
(apart
from a constant tangent-space Lorentz transformation)
$$e^a_\mu = \delta^a_\mu\eqno(52)$$
and a vanishing spin connection.
$$\omega = 0 \eqno(53)$$
Taking in (48c) the cosmological solution for $\eta_3$, allows solving (48e)
for $\eta_a$
$$\eta_a = {\Lambda\over 2}\epsilon_{a\mu} \left( x^\mu - x^\mu_0\right)
\eqno(54)$$
and from (48a) $\eta_2=2\eta$ is recovered to be as in (51).  Finally (49) is
solved for $a$.
$$a_\mu = {1\over 2} \epsilon_{\mu\nu}  x^\nu \eqno(55)$$
with a pure gauge contribution $\partial_\mu \chi$ left arbitrary.
The potential in (55)
corresponds to a constant ``magnetic field,'' as is appropriate with our
``magnetic-like'' extension of translations.

Note the invariants defined in (47): ${\cal F}^2$ vanishes since
$F^A$ does, $M$ is recognized as the ``black hole'' mass,
while $C$ is the cosmological constant.

The gauge theoretical solution may of course also be presented in a group
theoretical fashion, since the equations are of the same form as in (24) and
(25), with all quantities belonging to the {\it extended\/} algebra and group.
The explicit formulas,
corresponding to the ``black hole'' solution, Eqs.~(50) -- (55),
are as follows.  The group element $U$
that leads to the pure gauge connection $A=U^{-1} dU$ is
$$U = \exp x^aP_a\eqno(56)$$
up to a constant gauge transformation.
The constant algebra element $\Phi$ that gives
$H=\eta_a h^{ab} P_b - \eta_3 J - \eta_2 I = U^{-1} \Phi U$
is (placement of $\eta_2$ and $\eta_3$ dictated by the group metric
(42), {\it viz.\/} $\eta^A = h^{AB} \eta_B)$
$$\Phi = {\Lambda\over 2} x^a_0 \epsilon_a{}^b P_b
 + {\Lambda\over 2} J + \left( {M\over 2} - {\Lambda\over 4}
x_0^2 \right) I \eqno(57)$$

The above-mentioned gauge transformation can be used to set two of the
four constants $(x_0^a, M, \Lambda)$ to zero, leaving an invariant
dependence on the group scalars $M$ and $\Lambda$, i.e.~ on
$\eta^a \eta_a - 2 \eta_2 \eta_3$ and $i^A \eta_A$.

As in the (anti)~de~Sitter model, we see that after a further
gauge transformation we pass
to the geometrically singular configuration $A=0$, $H=\Phi$.  This gives an
especially succinct account of the relevant geometric information : $\Phi$
encodes the integration constants, which characterize the intrinsic geometry:
the cosmological constant $\Lambda$, the ``black hole'' mass $M$ and
location $x_0$. A geometry is
built with these characteristics once a gauge transformation is performed, say
with the above $U$, to obtain non-singular connections.
\goodbreak
\bigskip
\noindent{\bf IV.\quad QUANTIZATION}
\medskip
\nobreak
The gauge theoretical formulation
allows a succinct description
of the quantum theory.
Of course in the absence of matter,
which is all that we here consider,
there are no propogating degrees of freedom.
Nevertheless,
the quantal structure is interesting,
albeit simple.
\medskip
\line{\bf IV.1 \quad (Anti)~de~Sitter gravity. \hfil}
\smallskip
After a spatial integration by parts,
the Lagrange density (10) is given by
$$
{\cal L}''_1 =
\eta_{a} \dot{e}_1^a + \eta_2\dot{\omega}_1
+ e_0^a
\left(\eta'_a + \epsilon_a{}^b \eta_b\omega_1
-{\Lambda\over2}\eta_2 \epsilon_{ab} e_1^b \right)
+ \omega_0 \left( \eta'_2 + \eta_a \epsilon^a{}_b e_1^b \right) \eqno(58)
$$
Spatial end point contributions have been dropped, and the dot (dash)
denotes differentiation with respect to time (space).

The form (58) exhibits the canonical, symplectic structure, where
$(\eta_a, \eta_2)$ are canonical ``momenta'', $(e_a, \omega_1)$ are
conjugate ``coordinates'' and $(e_0^a, \omega_0)$ are Lagrange
multipliers, enforcing the vanishing of the constraints,
$$
\eqalignno{
G_a\ &=\ \eta'_a + \epsilon_a{}^b\eta_b\omega_1
- {\Lambda\over2}\eta_2\epsilon_{ab} e_1^b &(59a) \cr
G_2\ &=\ \eta'_2 + \eta_a \epsilon^a{}_b e_1^b &(59b) \cr }
$$
One readily verifies with the help of canonical commutation relations
between ``momenta'' and ``coordinates'' that the algebra of
constraints follows the Lie algebra (5).
$$
\eqalignno{
\left[ G_a (x), G_2 (y) \right]
\ &=\ i \epsilon_a{}^b\ G_b (x)\ \delta(x - y) \cr
\left[ G_a (x), G_b (y) \right]
\ &=\ -i{\Lambda\over2} \epsilon_{ab}\ G_2 (x)\ \delta(x - y) &(60)\cr}$$

The constraints annihilate physical states, which we take in a
Schr\"odinger, momentum representation; i.e.~states are functionals of
$(\eta_a, \eta_2)$ while $(e_a, \omega_1)$ act by functional
differentiation.  Thus physical states satisfy
$$ \left( \eta'_a + i \epsilon_a {}^b \eta_b {\partial\over{\partial\eta_2}}
-i{\Lambda\over2} \eta_2 \epsilon_{ab} {\partial\over{\partial\eta_b}} \right)
\!\Psi(\eta_a, \eta_2) = 0 \eqno(61a) $$
$$ \left( \eta'_2 + i \eta_a \epsilon_a {}^b
{\partial\over{\partial\eta_b}} \right)
\!\Psi(\eta_0, \eta_2) = 0 \eqno(61b) $$
The solution of these equations is
$$
\Psi(\eta_a, \eta_b) = \exp i\!\int dx
\left( \eta_2 \epsilon^{ab} \eta_a \eta'_b \Bigl/ \eta_c \eta^c \right)
\ \psi
\eqno(62)$$
where $\psi$ is a function (not functional) of the position-independent
part of the invariant $\eta^a \eta_a + {\Lambda\over2} \eta_2 \eta_2$.

\vfill\eject

\line{\bf IV.2 \quad Extended Poincar\'e gravity \hfil}
\smallskip
The Lagrangian density (44) is equivalently given by
$$\eqalignno{
{\cal L}_2{}'' &= \eta_a \dot{e}_1{}^a + \eta_2 \dot{\omega}_1
+ \eta_3 \dot{a}_1 + e_0^a
\left( \eta'_a + \epsilon_a {}^b \eta_b \omega_1
+ \eta_3 \epsilon_{a} {}_b e_1^b \right) \cr
&\quad{} + \omega_0
\left( \eta'_2 + \eta_a \epsilon^a {}_b e^b_1 \right)
+ a_0 \eta'_3 &(63)\cr}$$
This formula identifies the canonical conjugates
$(\eta_a, \eta_2, \eta_3)$
and
$(e_1^a, \omega_1, a_1)$
with
$(e_0^a, \omega_0, a_0)$
enforcing vanishing of the constraints
$$ \eqalignno{
G_a &= \eta'_a + \epsilon_a{}^b \eta_b \omega_1
+ \eta_3 \epsilon_a {}_b e_1^b &(64a)\cr
G_2 &= \eta'_2 + \eta_a \epsilon^a {}_b \ e_1^b &(64b)\cr
G_3 &= \eta'_3 &(64c)\cr } $$
whose commutators follow the Lie algebra (32)
$$ \eqalignno{
\left[ G_a (x), G_2 (y) \right]
\ &=\ i \epsilon_a {}^b\ G_b (x)\ \delta(x - y) \cr
\left[ G_a (x), G_b (y) \right]
\ &=\ i \epsilon_a {}_b\ G_3 (x)\ \delta(x - y) &(65) \cr} $$
Again we use a functional momentum representation
for the physical states, which must satisfy
$$ \left( \eta'_a + i \epsilon_a {}^b \eta_b {\partial\over{\partial\eta_2}}
+ i \eta_3 \epsilon_a {}_b {\partial\over{\partial\eta_b}} \right)
\!\Psi \left(\eta_a, \eta_2, \eta_3 \right) = 0 \eqno(65a) $$
$$ \left( \eta'_2 + i \eta_a \epsilon^a{}_b {\partial\over{\partial\eta_b}}
\right)
\!\Psi \left( \eta_a, \eta_2, \eta_3 \right) = 0 \eqno(65b) $$
$$ \eta'_3 \ \Psi \left( \eta_a, \eta_2, \eta_3 \right) = 0 \eqno(65c) $$

The general solution of these is
$$
\Psi \left( \eta_a, \eta_2, \eta_3 \right)
\ =\ \exp i \int
{dx \left( \eta_2 \epsilon^a {}^b \eta_a \eta'_b
\Bigl/ \eta_c \eta^c \right) \psi}
\eqno(66a)
$$
where $\psi$ depends on the constant part of the invariants
$\eta^c \eta_c - 2 \eta_2 \eta_3$
and $\eta_3 = -\eta_A i^A = -\Lambda / 2$.
These are essentially the black hole mass $M$
and the cosmological constant $\Lambda$.
Observe further that the phase in the exponent of (66a) may be written as
$$ {1\over2} \int {
{dx\over\eta_3}
\left( 2 \eta_2 \eta_3 - \eta^a \eta_a \right)
\epsilon^a {}^b \eta_a \eta'_b \Bigl/ \eta_c \eta^c}
\quad + \quad {1\over2}
\int { {dx\over\eta_3}\ \epsilon^a {}^b \eta_a \eta'_b }
\eqno(66b)  $$
But both $\eta_3$ and $\left( 2 \eta_2 \eta_3 - \eta^a \eta_a \right)$
contribute only their constant part and may be taken outside the integral.
The remaining integrand in the first integral is a total derivative, hence
that contribution to the phase may be ignored.
We are thus left with the wave functional
$$
\Psi\ =\ e^{-i\Lambda\!
\int\!{dx\,\epsilon^a {}^b \eta_a \eta'_b{} }}\;\psi(\Lambda,M)
\eqno(66c)
$$
Thus the cosmological constant and the black hole mass characterize a physical
state.
\goodbreak
\bigskip
\noindent{\bf IV.\quad CONCLUSION}
\medskip
\nobreak
The two models here considered are special: their geometric dynamics possess a
gauge theoretical formulation.  The extended Poincar\'e model exhibits
the intriguing possibility of a cosmological term that is an integration
constant, as are the ``black hole'' mass $M$ and location $x_0$; all three
are encoded in the Lagrange multipliers of the theory.

Both models can also be obtained by dimensional reduction from
$(2+1)$ dimensions:  To obtain (anti)~de~Sitter
gravity in its geometric formulation
one begins$^1$
with the Einstein theory/Hilbert action (with cosmological term),
suppresses dependence on the third dimension, sets $g_{\mu 2}$ to zero for
$\mu=0,1$ and $g_{22}$ to $\eta^2$; for the gauge theoretical formulation
one starts with the {\it Dreibein\/}-spin connection form of the theory, which
also is equivalent to a Chern--Simons, ${\cal O}(2,2)$ or ${\cal O}(3,1)$
model.$^5$  Extended Poincar\'e gravity can be similarly constructed, but the
higher-dimensional theory has to be suitably extended by an Abelian ideal.
Indeed it is found that {\it both\/} the (anti)~de~Sitter
and extended Poincar\'e
$(1+1)$ dimensional theories arise as {\it different\/} dimensional
reductions of a {\it single\/},
extended $(2+1)$-dimensional gravity.$^6$  This and another interesting topic
--- the coupling of matter consistently with the gauge principle$^7$ --- are
beyond the scope of our review.  In yet a further investigation one could
study non-topological theories in which invariants (46) and/or (47) are added
to the Lagrange density (43).

In conclusion, we note that dynamics determined by a group has been familiar
in physics since the invention of Yang--Mills theory.  However, the examples
described here offer a new possibility: in the Lie algebra that determines a
gauge theory one can allow an extension.  This gives rise to richer
dynamics within the same group theoretical structure, and in the gravity model
studied above produces the cosmological constant.
\goodbreak
\bigskip
\centerline{\bf ACKNOWLEDGEMENTS}
\medskip
\nobreak
The review was prepared with the assistance of D.~Cangemi, particularly in
finding the explicit solutions; this I gratefully acknowledge.
\vfill\eject

\centerline{\bf REFERENCES}
\medskip
\item{1.}C. Teitelboim, ``Gravitation and Hamiltonian Structure in Two
Space-Time Dimensions,'' {\it Phys. Lett.\/} {\bf 126B}, 41 (1983) and ``The
Hamiltonian Structure of Two-Dimensional Space-Time and its Relation with the
Conformal Anomaly,'' in {\it Quantum Theory of Gravity\/}, S. Christensen, ed.
(Adam Hilger, Bristol, 1984); R. Jackiw, ``Liouville Field Theory: A
Two-Dimensional Model for Gravity?,'' in {\it Quantum Theory of Gravity\/}, S.
Christensen, ed. (Adam Hilger, Bristol, 1984) and ``Lower-Dimensional
Gravity,'' {\it Nucl. Phys.\/} {\bf B252}, 343 (1985).
\medskip
\item{2.}H. Verlinde, ``Black Holes and Strings in Two Dimensions,'' in {\it
The Sixth Marcel Grossman Meeting\/}, H.~Sato
and T.~Nakamura, eds. (World Scientific, Singapore, 1992); C. Callan,
S. Giddings, A. Harvey and A.  Strominger, ``Evanescent Black Holes,''
{\it Phys. Rev. D\/} {\bf 45}, 1005 (1992).  Quantization of the model
is discussed in these papers.
\medskip
\item{3.}T. Fukuyama and K. Kamimura, ``Gauge Theory of Two-Dimensional
Gravity,'' {\it Phys. Lett.\/} {\bf 160B}, 259 (1985); K. Isler and C.
Trugenberger, ``Gauge Theory of Two-Dimensional Quantum Gravity,''
{\it Phys. Rev. Lett.\/} {\bf 63}, 834 (1989); A. Chamseddine
and D. Wyler, ``Gauge Theory of Topological Gravity in $1+1$ Dimensions,''
{\it Phys. Lett.\/} {\bf B228}, 75 (1989).  Quantization of the model is
discussed in these papers.
\medskip
\item{4.}D. Cangemi and R. Jackiw, ``Gauge Invariant Formulations of Lineal
Gravity,'' {\it Phys. Rev. Lett.\/} {\bf 69}, 233 (1992).
\medskip
\item{5.}A.~Ach\'ucarro and P.~Townsend, ``A Chern--Simons Action for
Three-Dimensional anti-de~Sitter Supergravity Theories,'' {\it
Phys.~Lett.\/} {\bf B180}, 89 (1986); E. Witten, ``$2+1$-Dimensional
Gravity as an Exactly Solvable System,'' {\it Nucl.~Phys.\/} {\bf
B311}, 46 (1988/89).
\medskip
\item{6.}D. Cangemi, ``One Formulation for both Lineal Gravities
through a Dimensional Reduction,'' {\it Phys. Lett.} {\bf B297}, 261
(1992); A.~Ach\'ucarro, ``Lineal Gravity from Planar Gravity,'' {\it
Phys.~Rev.~Lett.} {\bf 70}, 1037 (1993).
\medskip
\item{7.}G. Grignani and G. Nardelli,
``Poincar\'e Gauge Theories for Lineal Gravities,''
{\it Nucl.~Phys.\/}~{\bf B} (in press);
D.~Cangemi and R. Jackiw, ``Geometric
Gravitational Force on Particles Moving in a Line,'' {\it Phys.~Lett.}
{\bf B229}, 24 (1993) and ``Poincar\'e Gauge Theory for Gravitational
Forces in (1+1) Dimensions,'' {\it Ann.~Phys.} (NY) {\bf 225}, 229 (1993).

\bye